# $Al_4SiC_4$ würtzite crystal: structural, optoelectronic, elastic and piezoelectric properties

L. Pedesseau,[1,a)] J. Even,[1,a)] M. Modreanu,[2] D. Chaussende,[3] O. Chaix-Pluchery[3] and O. Durand,[1]

[1]*UMR FOTON, CNRS, INSA Rennes, Rennes, F35708, France*

[2]*Tyndall National Institute, University College Cork, Lee Maltings, Cork, Ireland*

[3]*Univ. Grenoble Alpes, LMGP, 38000 Grenoble, France and CNRS, LMGP, 38000 Grenoble, France*

New experimental results supported by theoretical analyses are proposed for aluminum silicon carbide ($Al_4SiC_4$). A state of the art implementation of the Density Functional Theory is used to analyze the experimental crystal structure, the Born charges, the elastic and piezoelectric properties. The Born charge tensor is correlated to the local bonding environment for each atom. The electronic band structure is computed including self-consistent many-body corrections. $Al_4SiC_4$ material properties are compared to other wide band gap Würtzite materials. From a comparison between an ellipsometry study of the optical properties and theoretical results, we conclude that the $Al_4SiC_4$ material has indirect and direct band gap energies of about 2.5eV and 3.2 eV respectively.

Pursuing competitive and sustainable solutions for next generation of electronic, photonic and solar technologies, the manufacturers seek low cost, abundant and non-toxic materials. Aluminum silicon carbide alloys are poised to set off such innovations, leading potentially to transparent conductive oxides (TCO), aluminum silicon carbon based LED, wide band gap optoelectronics, attractive mechanical properties and high temperature operation for electronics.

$Al_4SiC_4$ is a low cost potential material for refractory and high temperature applications, receiving attention due to its low density (3.03 g/cm$^3$), in comparison with traditional thermal materials such as Cu - CuMo - CuW (9 g/cm$^3$ - 10 g/cm$^3$ - 16 g/cm$^3$), high melting point (~2353 K) and excellent oxidation resistance[1–3]. High temperature behavior was investigated a few years ago through thermal conductivity[4] and electrical resistivity[5,6] measurements. The impact of oxidation on the electrical resistance was also studied[7]. $Al_4SiC_4$ is one of the polymorphs in the aluminum silicon carbide family[1] that can be found in ceramic materials. This material has various potential[8] applications in aerospace and automotive industries, owing to its superior strength to weight ratio and high temperature resistance. Its thermal expansion coefficient is also suited for direct integrated circuit device attachment with a maximum thermal dissipation of about 200 W/mK. In addition, this material can also be used to protect hermitically the sensitive electronic components over the environment.

[a)] Authors to whom correspondence should be addressed. Electronic mail: laurent.pedesseau@insa-rennes.fr and jacky.even@insa-rennes.fr .

$Al_4SiC_4$ is known for almost six decades since early optical and X-ray diffraction characterizations[9]. This material crystallizes in a yellow form belonging to the hexagonal system. The bonding characteristics, elastic stiffness, ideal strengths, and atomistic deformation modes of $Al_4SiC_4$ were investigated by using Density functional theory (DFT) theory [10]. The electronic band structure and optical properties were simulated at the same level of theory[11] , the $Al_4SiC_4$ crystal being predicted to be a small gap semiconductor with an $E_g^{\Gamma \to M}$ indirect band gap of 1.05 eV. No experimental results on the optical properties were available up to now to confirm these predictions.

In this paper, we perform an experimental and theoretical study of the structural, electronic, optical, elastic and piezoelectric properties of the $Al_4SiC_4$ crystal. The crystal structure simulated at the DFT level is compared to available X-ray diffraction data[1]. Density functional perturbation theory (DFPT) is used to study the Born dynamical effective charges, elastic and piezoelectric properties of the $Al_4SiC_4$ crystal. The electronic band structure is calculated including self-consistent many-body (scGW) contributions [12,13] to correct the well-known underestimation of the band gaps computed DFT level. The electronic band structure is used to analyse new optical data from phase modulated spectroscopic ellipsometry.

DFT calculations are performed using the plane-wave projector augmented wave (PAW) method as implemented in the VASP code[14–16]. The local density approximation (LDA) is used for the exchange-correlation functional[17]. Fully dynamical scGW[18–20] many-body corrections are included to compute the electronic band structure and monoelectronic state wavefunctions. Norm-conserving pseudopotentials are constructed for Al [$3s^2 3p^1$], Si [$3s^2 3p^2$] and C [$2s^2 2p^2$] atoms. Atomic valence configurations are similar to the ones used in previous DFT simulations of the ground state[10,11]. A plane-wave basis set with an energy cut-off of 950 eV is used to expand the electronic wave-functions. The reciprocal space integration is performed over a 18x18x3 Monkhorst-Pack grid[21]. Energy convergence is accurately reached with tolerance on the residual potential which stems from difference between the input and output potentials. The crystal structure has been relaxed until the forces acting on each atom are smaller than $10^{-6}$eV/Å.

$Al_4SiC_4$ single crystals have been obtained by high temperature synthesis, high purity silicon (9N) and aluminium (99.5) pieces were melted in a graphite crucible which acted both as a container for the melt and as carbon source. $Al_4SiC_4$ single crystals were grown by maintaining the melt at high temperature (1800$^o$C) before cooling down at a very low and controlled rate. The crystal structure was confirmed by TEM and XRD to be Al4SiC4 phase in hexagonal structure (Space group $P6_3mc$) with lattice parameters a=0.32812±0.00045nm and c=2.17042±0.00554[22].

The optical properties were investigated using phase modulated spectroscopic Ellipsometry (UVISEL-Jobin Yvon). Ellipsometry yields the ratio, ρ, of the Fresnel reflection coefficient of the p-polarized (parallel to the plane of incidence of the linearly polarized light beam) and s-polarized (perpendicular to the plane of incidence) light reflected from the surface through the Ellipsometry angles Ψ and Δ defined by the equation:

$$\rho = \tan(\Psi)\exp(i\Delta)$$

ρ and, hence, Ψ and Δ, are related to the material pseudodielectric function, $\langle\varepsilon\rangle = \langle\varepsilon_1\rangle + i\langle\varepsilon_2\rangle$, through the equation:

$$\langle\varepsilon\rangle = \sin\phi^2\left[1+\tan^2\phi\frac{(1-\rho)^2}{(1+\rho)^2}\right]$$ where $\phi$ is the angle incidence.

The Spectroscopic Ellipsometry data (Ψ,Δ) have been measured on the $Al_4SiC_4$ compound between 1eV and 5eV at $70^o$ angle of incidence. The pseudodielectric function for $Al_4SiC_4$ has been deduced from the experimental data using a simple two-phase model consisting of $Al_4SiC_4$/air.

- Crystal structure

We have performed a full optimization of the hexagonal structure (space group $P6_3mc$) by minimizing the total energy with respect to the lattice constants a and c and the internal positions of each atoms in the unit cell. Figure 1 shows an overview of the crystal structure projected on the (a,c) and (a,b) planes. The DFT results agree rather well with experimental data[1] as reported in Table I : discrepancies are found less than 2%. The calculated ratio c/a=6.630[10], c/a=6.646[11], c/a=6.619 and c/a=6.610 respectively obtained at GGA ultrasoft, GGA, LDA and PAW LDA levels, differ only slightly from the experiment value 6.614.

- Born dynamical effective charges, Elastic and Piezoelectric constants

The knowledge of the local bonding structures and local charges allows investigating the correlation between these quantities. The Born dynamical effective charge tensor[23] is calculated by the second derivatives of the total energy E with respect to one atomic displacement u and one component of electric field. This tensor is then used to calculate the atomic relaxation contribution to both the elastic and the piezoelectric tensors.

To correlate the Born effective charges with the local bonding structures, nine building blocks have been isolated (Fig 2.): one $SiC_4$ block, four $AlC_4$ blocks, and also four other blocks each containing a carbon atom e.g. $CSi_3Al$, $CSiAl_4$, $CAl_5$ and $CAl_6$ from the $Al_4SiC_4$ material. On Figure 2, the Bond lengths and bond Angles in each block have been drawn, with atomic notations according to Z. Inoue[1]. The $SiC_4$ block and the four $AlC_4$ blocks are found to be distorted tetrahedral structures, which is is consistent with previous results[11]. The four last blocks are completely different from the first 5 ones. In $CSi_3Al$, the carbon atom (C4) is in coordination four which corresponds to a tetrahedral geometry. In $CSiAl_4$ and $CAl_{5,}$ the carbon atoms (C2 and C3) are in coordination five which corresponds to a trigonal bipyramid geometry. In particular, the carbon atom C2 is bridging the $SiC_4$ and $AlC_4$ tetrahedral blocks. Finally, the Carbon atom C1 is in coordination six in $CAl_6$ and is located in a middle of a trigonal prism geometry. The carbon atoms C1 and C3 are in the middle of the $AlC_4$ tetrahedral entities. The variety of local environments can be characterized by the mean value of the Born charge $Z_{iso}^{X}$, computed from the trace of the Born charge tensor (Table II). These charges are significantly different from the nominal valence charge (*i.e.* +3 for Al, +4 for Si, +4 for C), which indicates a mixed ionic-covalent nature of the interactions. For the Aluminum atoms, the standard deviation is small (0.1) with an average of $Z_{iso}^{Al}$ equal to 2.50 and the $Z_{iso}^{Si}$ charge of the Silicon is equal to about 2.77. Finally, the carbon atom case is more complicated due to the different local coordinations. The absolute value of $Z_{iso}^{C}$ is increasing as function of the coordination number, from -2.93 for a coordination equal to 4, to -3.65 for a coordination equal to 6.

The $Al_4SiC_4$ Würtzite crystal has five independent elastic constants (Table III), namely, $C_{11}$, $C_{12}$, $C_{13}$, $C_{33}$ and $C_{44}$, defined in Cartesian coordinates, assuming a symmetry axis along z. Due to the hexagonal symmetry, the $C_{66}$ constant depends on $C_{11}$ and $C_{12}$: $C_{66}=(C_{11}-C_{12})/2$. In order to estimate the elastic constants, the second-order strain derivatives of total energy were computed within DFPT[23]. Table III shows that the results for the elastic constants and bulk modulus $B_o$, are in good agreement with other DFT results from the litterature[10] and experimental data[24]. The $Al_4SiC_4$ material can be compared to other common wide band gap Würtzite materials: the bulk modulus of $Al_4SiC_4$ is higher by about 40GPa than the ones of ZnO[25] (143.6GPa) and InN[26] (140.9GPa) materials and smaller by about 30-40GPa than the ones of GaN[27] (210.2GPa), AlN[26] (207.9GPa) and SiC[28] (220.6GPa) materials.

To obtain the piezoelectric coefficients, the second-order derivatives of total energy have been calculated with respect to the strain and the electric field components[23] by DFPT. Small values are predicted for the linear piezoelectric coefficients (Tab. III): $e_{15}=-0.04$ C.m$^{-2}$, $e_{33}=+0.36$ C.m$^{-2}$, $e_{31}=+0.03$ C.m$^{-2}$ by comparison with conventional semiconductors in zinc-blende[29–32] or Würtzite structures[33–35] using similar procedures. Indeed, these constants are one order of magnitude smaller than the ones of AlN, GaN, InN and ZnO semiconductors[35] in Würtzite structure. The piezoelectric constants are nevertheless close to the ones of SiC[36] e.g. $e_{15}$=+0.08 $C.m^{-2}$, $e_{33}$=+0.20 $C.m^{-2}$. In order to understand the origin of the piezoelectric response in $Al_4SiC_4$, and to explain its small values in comparison with most Würtzite semiconductors, the proper piezoelectric component $e_{ik}$ has been decomposed into an external $e_{ik}^{0}$ and internal $e_{ik}^{int}$ contributions[37–39]. The contributions are reported in Table III. The external and internal contributions are almost compensated from one to each other, for both $e_{15}$ and $e_{31}$ components leading to small values for these linear piezoelectric coefficients.

The experimental variations of the real and imaginary parts of the dielectric constants are reported in Fig. 3. The experimental variation of the imaginary part exhibits a small peak around 2.3eV, which can be likely attributed to SiC impurities, remaining in the polycrystalline $Al_4SiC_4$ material after the growth procedure. SiC has indeed about one hundred polytypes[40] with gap energies $E_g^{SiC}$ in the $2.2eV \leq E_g^{SiC} \leq 3.3$ range. The absorption of the polycrystalline $Al_4SiC_4$ material starts to increase significantly from energy values larger than 2.5 eV. Two contributions can be distinguished, a first one starting at 2.5 eV and a second one at 3.2 eV respectively attributed to indirect and direct electronic band to band transitions.

In order to interpret the optical measurements, a first computation of the electronic band structure was performed exactly at the same DFT theory level than the previous work with the same k-grid (4x4x1) and plane-wave cutoff of 500eV[11]. The results are essentially similar to the ones reported in the left column of table IV. The computed band gaps, even the direct ones, are clearly too small to explain the onset of the experimental optical absorption located at about 3.2eV. Next results were obtained with enhanced numerical conditions including a PAW decomposition, an extended k-grid (18x18x3) and a plane-wave cutoff of 950eV. Using adapted PAW potentials, available with the VASP code, the computed indirect gap energy $E_g^{\Gamma\rightarrow M}$ was increased by about 0.9eV, reaching a value of 2eV. Finally, self-consistent GW many-body corrections have been added to yield a more realistic electronic band structure (Figure 4). The simulated indirect band gap $E_g^{\Gamma\rightarrow M}$ now reaches a value of 2.48 eV in good agreement with the first absorption contribution (Figure 3). Furthermore, the first direct gap energy is equal to about 3.2eV $E_g^{M\,to\,L}$ which corresponds to the onset of the second significant contribution to the absorption (figure 3). We thus conclude that the $Al_4SiC_4$ material displays indirect and direct band gap energies of about 2.5eV and 3.2 eV respectively. These values are similar to the band gap values of most Würtzite Wide band gap materials such as ZnO[41] (3.44eV), GaN[42] (3.30eV), AlN[42] (5.4eV) and SiC[43] (3.33eV).

In summary, this paper presents an analysis of the structural properties of the wide bandgap $Al_4SiC_4$ wurtzite compound, including the crystal structure, Born effective charges, Elastic and Piezoelectric properties. The piezoelectric constants are reported for the first time. The electronic band structure is computed using DFT plus scGW corrections and used to interpret new experimental data on the optical properties. From a comparison between the experimental absorption spectrum and theoretical results, it is found that the $Al_4SiC_4$ material has indirect and direct band gap energies of about 2.5eV and 3.2 eV respectively.

## ACKNOWLEDGMENTS

The *ab initio* simulations have been performed on HPC resources of CINES under the allocation 2015-[x2015096724] made by GENCI (Grand Equipement National de Calcul Intensif).

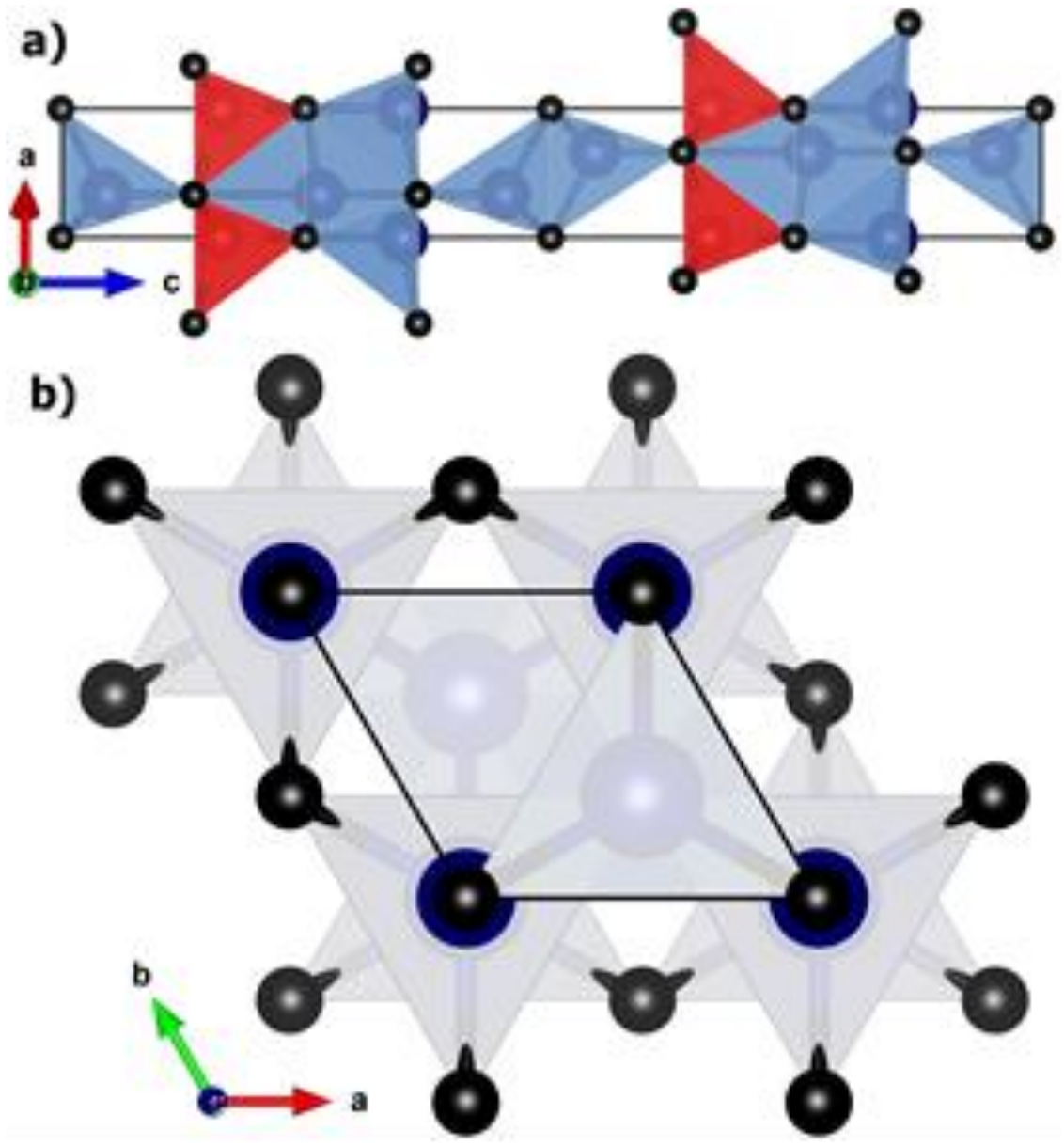

FIG. 1. (Color online) Overview of the crystal structure of $Al_4SiC_4$ projected on the ac (a) and ab (b) planes (Black: Carbon; Blue: Aluminum; Red: Silicon)

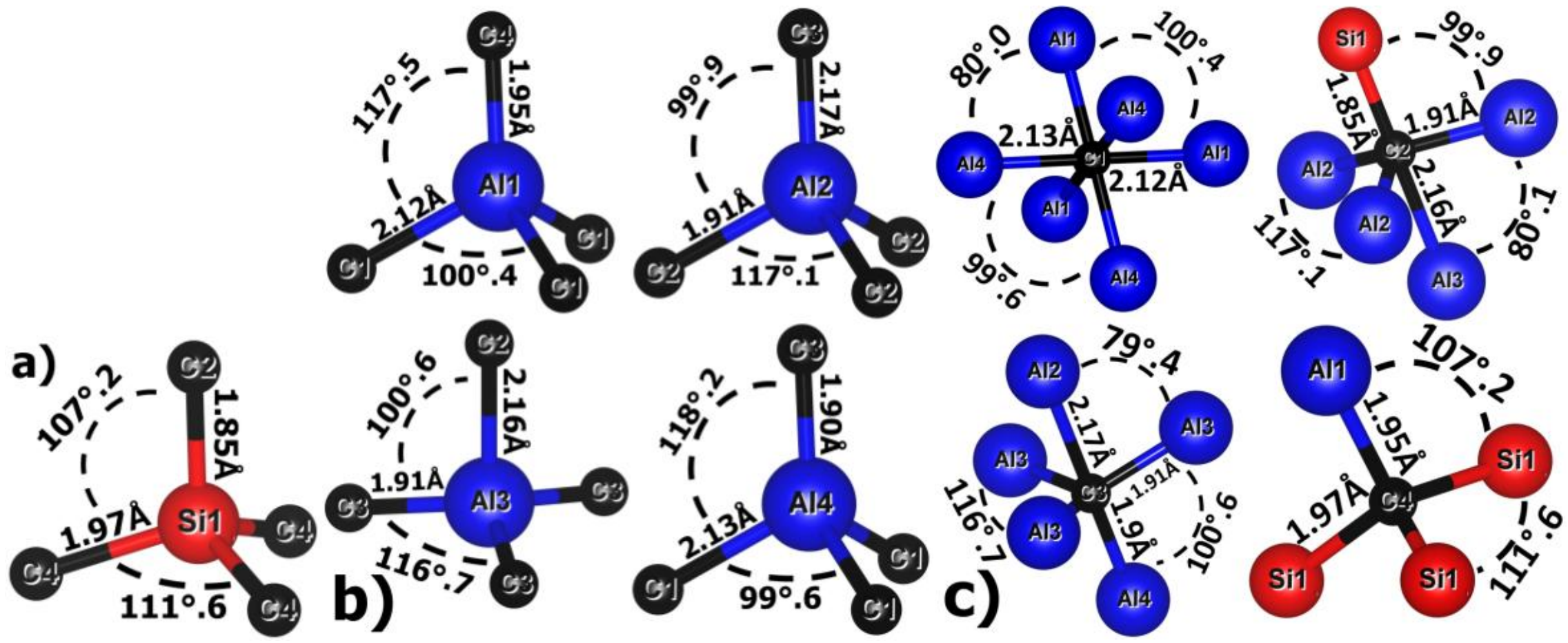

FIG. 2. (Color online) Bond lengths in Angstrom and bond Angles in degree of each entity namely (a) $SiC_4$, (b) $AlC_4$ and (c) $CSi_3Al$, $CSi_1Al_4$, $CAl_5$ and $CAl_6$ of $Al_4SiC_4$. The atomic notations come from Z. Inoue[1]. (Black: Carbon; Blue: Aluminum; Red: Silicon)

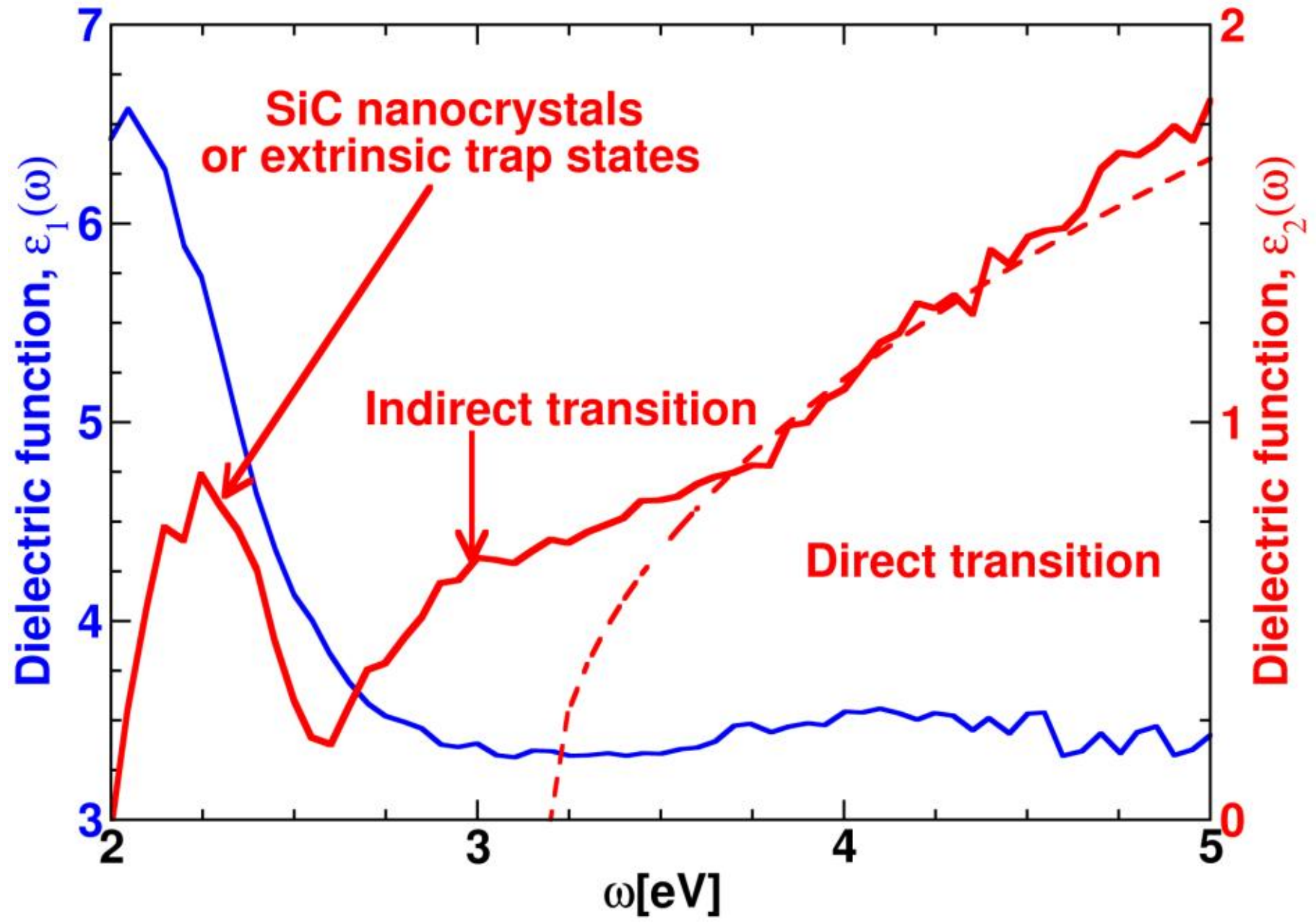

FIG. 3. (Color online) Real (blue) and Imaginary (red) parts of the Dielectric function of the $Al_4SiC_4$ crystal structure. Dash curve (red) is the imaginary part of the Dielectric function fitted for the direct transition.

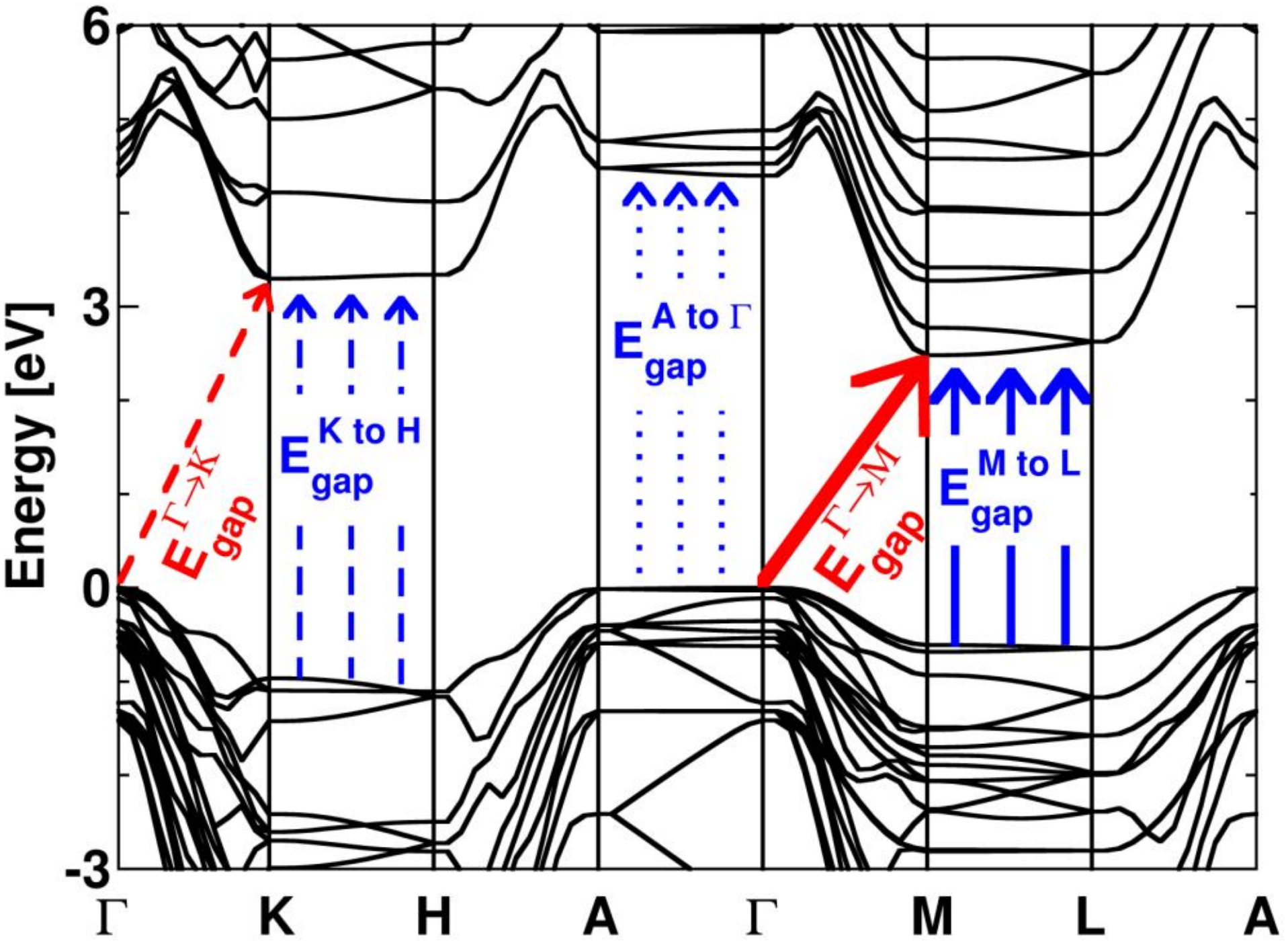

FIG. 4. (Color online) DFT electronic band structures of $Al_4SiC_4$ including many-body corrections. Energy of the valence band maximum is set at zero. $\mathbf{E}_{\mathbf{gap}}^{\Gamma\to\mathbf{M}}$ and $\mathbf{E}_{\mathbf{gap}}^{\Gamma\to\mathbf{K}}$ are indirect band gaps, and $\mathbf{E}_{\mathbf{gap}}^{\mathbf{M\ to\ L}}$, $\mathbf{E}_{\mathbf{gap}}^{\mathbf{K\ to\ H}}$, $\mathbf{E}_{\mathbf{gap}}^{\mathbf{A\ to\ \Gamma}}$ direct band gaps (red tilted arrows for indirect energy gaps; blue vertical arrows for direct energy gaps)

TABLE I. DFT calculations compared to other simulations[10,11] and experimental data[1,22] of the lattice parameters in Å of the $Al_4SiC_4$.

| | | *Other DFT*[10] *(GGA)* | *Other DFT*[11] *(GGA)* | *DFT (LDA)* | *DFT (PAW LDA)* | *Expt.*[22] | *Other expt.*[1] |
|---|---|---|---|---|---|---|---|
| Lattice parameters [Å] | a=b | 3.22(1.8%) | 3.28(0.1%) | 3.23(1.5%) | 3.26 (0.6%) | 3.28 | 3.28 |
| | c | 21.35(1.5%) | 21.80(0.5%) | 21.38 (1.4%) | 21.55(0.6%) | 21.70 | 21.68 |

TABLE II. DFT calculations of the $Z_{ij}$ (and with $Z_{iso}$=Trace(Z)/3) Born charge tensors for each element of the $Al_4SiC_4$.

| Atom Type | WP | $Z_{XX}$ | $Z_{YY}$ | $Z_{ZZ}$ | $Z_{ij}$ with i≠j | $Z_{iso}$ |
|---|---|---|---|---|---|---|
| Si1 | 2a | 2.72 | 2.72 | 2.86 | 0.0 | 2.77 |
| Al1 | 2b | 2.53 | 2.53 | 2.71 | 0.0 | 2.59 |
| Al2 | 2b | 2.22 | 2.22 | 2.86 | 0.0 | 2.43 |
| Al3 | 2b | 2.29 | 2.29 | 2.49 | 0.0 | 2.36 |
| Al4 | 2b | 2.52 | 2.52 | 2.79 | 0.0 | 2.61 |
| C1 | 2a | -3.84 | -3.84 | -3.26 | 0.0 | -3.65 |
| C2 | 2a | -2.67 | -2.67 | -3.97 | 0.0 | -3.10 |
| C3 | 2b | -2.83 | -2.83 | -3.57 | 0.0 | -3.08 |
| C4 | 2b | -2.94 | -2.94 | -2.91 | 0.0 | -2.93 |

TABLE III. DFT calculations compared to a previous simulation[10] of the elastic constants in GPa, the Bulk modulus in GPa, and piezoelectric constants, external and internal contributions in C/m² of the $Al_4SiC_4$. The experimental data are given in parentheses [24].

| | | *Other DFT*[10] | *This study* |
|---|---|---|---|
| Elastic constants [GPa] | $C_{11}$ | 386 | 383.6 |
| | $C_{12}$ | 118 | 121.9 |
| | $C_{13}$ | 50 | 51.1 |
| | $C_{33}$ | 409 | 411.0 |
| | $C_{44}$ | 122 | 118.4 |
| | $C_{66}$ | 134 | 130.8 |
| Bulk modulus [GPa] | $B_o$ | 179 | 180.3 (182[24]) |
| Piezoelectric constants [C/m²] | $e_{15}$ | | -0.04 |
| | $e_{33}$ | | 0.36 |
| | $e_{31}$ | | 0.03 |
| External Piezoelectric contributions [C/m²] | $e_{15}^{0}$ | | 0.13 |
| | $e_{33}^{0}$ | | -0.11 |
| | $e_{31}^{0}$ | | 0.10 |
| Internal Piezoelectric contributions [C/m²] | $e_{15}^{int}$ | | -0.17 |
| | $e_{33}^{int}$ | | 0.47 |
| | $e_{31}^{int}$ | | -0.07 |

TABLE IV. Previous DFT (GGA)[11], and DFT (PAW-LDA)+scGW calculations of the direct and indirect electronic energy gaps along the Γ–A, K–H, Γ–K, M–L, Γ–M directions for the $Al_4SiC_4$ crystal.

| | *Previous DFT study* [11] | *DFT+scGW* |
|---|---|---|
| $E_g^{\Gamma\,to\,A}$ direct [eV] (5th) | $3.30eV \leq E_g^{\Gamma\,to\,A} \leq 3.36eV$ | $4.40eV \leq E_g^{\Gamma\,to\,A} \leq 4.48eV$ |
| $E_g^{K\,to\,H}$ direct [eV] (4th) | $3.80eV \leq E_g^{K\,to\,H} \leq 3.91eV$ | $4.26eV \leq E_g^{K\,to\,H} \leq 4.45eV$ |
| $E_g^{\Gamma\rightarrow K}$ indirect [eV] (3rd) | 2.02eV | 3.30eV |
| $E_g^{M\,to\,L}$ direct [eV] (2nd) | $1.91eV \leq E_g^{M\,to\,L} \leq 2.01eV$ | $3.10eV \leq E_g^{M\,to\,L} \leq 3.25eV$ |
| **$E_g^{\Gamma\rightarrow M}$ indirect [eV] (1rt)** | **1.05** | **2.48** |

## REFERENCES

1 Z. Inoue, Y. Inomata, H. Tanaka, and H. Kawabata, J. Mater. Sci. **15**, 575 (1980).
2 K. Inoue, S. Mori, and A. Yamaguchi, J. Ceram. Soc. Jpn. **111**, 348 (2003).
3 K. Inoue and A. Yamaguchi, J. Am. Ceram. Soc. **86**, 1028 (2003).
4 K. Inoue, S. Mori, and A. Yamaguchi, J. Ceram. Soc. Jpn. **111**, 466 (2003).
5 K. Inoue and A. Yamaguchi, J. Ceram. Soc. Jpn. **111**, 267 (2003).
6 K. Inoue, A. Yamaguchi, and S. Hashimoto, J. Ceram. Soc. Jpn. **110**, 1010 (2002).
7 K. Inoue, S. Mori, and A. Yamaguchi, J. Ceram. Soc. Jpn. **111**, 126 (2003).
8 L.D. Hefti, J. Mater. Eng. Perform. **16**, 136 (2007).
9 V.J. Barczak, J. Am. Ceram. Soc. **44**, 299 (1961).
10 T. Liao, J. Wang, and Y. Zhou, Phys. Rev. B **74**, 174112 (2006).
11 A. Hussain, S. Aryal, P. Rulis, M. Choudhry, and W. Ching, Phys. Rev. B **78**, 195102 (2008).
12 L. Hedin, Phys. Rev. **139**, A796 (1965).
13 M. van Schilfgaarde, T. Kotani, and S. Faleev, Phys. Rev. Lett. **96**, 226402 (2006).

[14] G. Kresse and J. Furthmüller, Phys. Rev. B **54**, 11169 (1996).
[15] G. Kresse and J. Furthmüller, Comput. Mater. Sci. **6**, 15 (1996).
[16] M. Shishkin and G. Kresse, Phys. Rev. B **74**, 035101 (2006).
[17] M. Fuchs and M. Scheffler, Comput. Phys. Commun. **119**, 67 (1999).
[18] M. Shishkin and G. Kresse, Phys. Rev. B **75**, 235102 (2007).
[19] F. Fuchs, J. Furthmüller, F. Bechstedt, M. Shishkin, and G. Kresse, Phys. Rev. B **76**, 115109 (2007).
[20] M. Shishkin, M. Marsman, and G. Kresse, Phys. Rev. Lett. **99**, 246403 (2007).
[21] J. Pack and H. Monkhorst, Phys. Rev. B **16**, 1748 (1977).
[22] D. Zevgitis, O. Chaix-Pluchery, B. Doisneau, M. Modreanu, J. La Manna, E. Sarigiannidou, and D. Chaussende, Mater. Sci. Forum **821-823**, 974 (2015).
[23] X. Wu, D. Vanderbilt, and D.R. Hamann, Phys. Rev. B **72**, 035105 (2005).
[24] V.L. Solozhenko and O.O. Kurakevych, Solid State Commun. **135**, 87 (2005).
[25] T.B. Bateman, J. Appl. Phys. **33**, 3309 (1962).
[26] A.F. Wright, J. Appl. Phys. **82**, 2833 (1997).
[27] A. Polian, M. Grimsditch, and I. Grzegory, J. Appl. Phys. **79**, 3343 (1996).
[28] K. Kamitani, M. Grimsditch, J.C. Nipko, C.-K. Loong, M. Okada, and I. Kimura, J. Appl. Phys. **82**, 3152 (1997).
[29] G. Bester, X. Wu, D. Vanderbilt, and A. Zunger, Phys. Rev. Lett. **96**, 187602 (2006).
[30] G. Bester, A. Zunger, X. Wu, and D. Vanderbilt, Phys. Rev. B **74**, 081305 (2006).
[31] J. Even, F. Doré, C. Cornet, L. Pedesseau, A. Schliwa, and D. Bimberg, Appl. Phys. Lett. **91**, 122112 (2007).
[32] L. Pedesseau, J. Even, F. Doré, and C. Cornet, in *AIP Conf. Proc.* (AIP Publishing, 2007), pp. 1331–1334.
[33] L. Pedesseau, C. Katan, and J. Even, Appl. Phys. Lett. **100**, 031903 (2012).
[34] A. Beya-Wakata, P.-Y. Prodhomme, and G. Bester, Phys. Rev. B **84**, 195207 (2011).
[35] P.-Y. Prodhomme, A. Beya-Wakata, and G. Bester, Phys. Rev. B **88**, 121304 (2013).
[36] Y. Goldberg, M.E. Levinshtein, and S.L. Rumyantsev, in *Prop. Adv. Semicond. Mater. GaN AlN SiC BN SiC SiGe*, John Wiley & Sons, Inc., New York (Eds. Levinshtein M.E., Rumyantsev S.L., Shur M.S., 2001), pp. 93–148.
[37] A. Dal Corso, M. Posternak, R. Resta, and A. Baldereschi, Phys. Rev. B **50**, 10715 (1994).
[38] M. Catti, Y. Noel, and R. Dovesi, J. Phys. Chem. Solids **64**, 2183 (2003).
[39] D. Vanderbilt, J. Phys. Chem. Solids **61**, 147 (2000).
[40] R. Cheung, *Silicon Carbide Microelectromechanical Systems for Harsh Environments* (Imperial College Press, 2006).
[41] S.F. Chichibu, T. Sota, G. Cantwell, D.B. Eason, and C.W. Litton, J. Appl. Phys. **93**, 756 (2003).
[42] I. Vurgaftman and J.R. Meyer, J. Appl. Phys. **94**, 3675 (2003).
[43] L. Patrick, D.R. Hamilton, and W.J. Choyke, Phys. Rev. **143**, 526 (1966).